\PassOptionsToPackage{unicode}{hyperref}
\PassOptionsToPackage{hyphens}{url}
\documentclass[
  11pt,
]{article}
\usepackage{xcolor}
\usepackage[margin=1in]{geometry}
\usepackage{amsmath,amssymb}
\usepackage{iftex}
\ifPDFTeX
  \usepackage[T1]{fontenc}
  \usepackage[utf8]{inputenc}
  \usepackage{textcomp} 
\else 
  \usepackage{unicode-math} 
  \defaultfontfeatures{Scale=MatchLowercase}
  \defaultfontfeatures[\rmfamily]{Ligatures=TeX,Scale=1}
\fi
\usepackage{lmodern}
\ifPDFTeX\else
\fi
\IfFileExists{upquote.sty}{\usepackage{upquote}}{}
\IfFileExists{microtype.sty}{
  \usepackage[]{microtype}
  \UseMicrotypeSet[protrusion]{basicmath} 
}{}
\makeatletter
\@ifundefined{KOMAClassName}{
  \IfFileExists{parskip.sty}{%
    \usepackage{parskip}
  }{
    \setlength{\parindent}{0pt}
    \setlength{\parskip}{6pt plus 2pt minus 1pt}}
}{
  \KOMAoptions{parskip=half}}
\makeatother
\usepackage{color}
\usepackage{fancyvrb}

\DefineVerbatimEnvironment{Highlighting}{Verbatim}{commandchars=\\\{\}}
\newenvironment{Shaded}{}{}

\newcommand{\BuiltInTok}[1]{\textcolor[rgb]{0.00,0.50,0.00}{#1}}

\newcommand{\ControlFlowTok}[1]{\textcolor[rgb]{0.00,0.44,0.13}{\textbf{#1}}}

\newcommand{\DecValTok}[1]{\textcolor[rgb]{0.25,0.63,0.44}{#1}}

\newcommand{\FloatTok}[1]{\textcolor[rgb]{0.25,0.63,0.44}{#1}}

\newcommand{\KeywordTok}[1]{\textcolor[rgb]{0.00,0.44,0.13}{\textbf{#1}}}
\newcommand{\NormalTok}[1]{#1}
\newcommand{\OperatorTok}[1]{\textcolor[rgb]{0.40,0.40,0.40}{#1}}

\newcommand{\StringTok}[1]{\textcolor[rgb]{0.25,0.44,0.63}{#1}}

\usepackage{longtable,booktabs,array}
\usepackage{float}
\usepackage{adjustbox}
\usepackage{calc} 
\usepackage{etoolbox}
\makeatletter
\patchcmd\longtable{\par}{\if@noskipsec\mbox{}\fi\par}{}{}
\makeatother
\IfFileExists{footnotehyper.sty}{\usepackage{footnotehyper}}{\usepackage{footnote}}
\makesavenoteenv{longtable}
\providecommand{\tightlist}{%
  \setlength{\itemsep}{0pt}\setlength{\parskip}{0pt}}
\usepackage{bookmark}
\IfFileExists{xurl.sty}{\usepackage{xurl}}{} 
\hypersetup{
  hidelinks,
  pdfcreator={LaTeX via pandoc}}

\title{Privacy-Aware Split Inference with Speculative Decoding for Large Language Models over Wide-Area Networks}
\author{Michael Cunningham}
\date{February 2026}

\begin{document}

\maketitle

\begin{abstract}

We present a practical system for privacy-aware large language model
(LLM) inference that splits a transformer between a trusted local GPU
and an untrusted cloud GPU, communicating only intermediate activations
over the network. Our system addresses the unique challenges of
autoregressive LLM decoding over high-latency wide-area networks (WANs),
contributing: (1) an asymmetric layer split where embedding and
unembedding layers remain local, ensuring raw tokens never leave the
trusted device; (2) the first application of lookahead decoding to split
inference over WANs, amortizing network round-trip latency across
multiple tokens per iteration; (3) an empirical inversion attack
evaluation showing that split depth provides a tunable
privacy-performance tradeoff---an attacker can recover
\textasciitilde59\% of tokens at a 2-layer split but only
\textasciitilde35\% at an 8-layer split, with minimal throughput impact;
(4) ablation experiments showing that n-gram speculation accepts
1.2--1.3 tokens per decoding step on average (peak of 7 observed on
code), with acceptance rates consistent across model scales; (5) formal
verification that lookahead decoding produces token-identical output to
sequential decoding under greedy argmax, with zero quality degradation;
and (6) scaling validation on Mistral NeMo 12B (40 layers),
demonstrating that the system generalizes to larger models with only 4.9
GB local VRAM and matching 7B throughput. Evaluated on Mistral 7B and
NeMo 12B over a \textasciitilde80ms WAN link, our system achieves
8.7--9.3 tok/s (7B) and 7.8--8.7 tok/s (12B) with lookahead decoding,
with an RTT decomposition model (validated at \textless6.2\%
cross-validation error) projecting 15--19 tok/s at 20ms RTT.
\end{abstract}

\section{Introduction}\label{sec:introduction}

The rapid adoption of large language models in enterprise settings has
created a fundamental tension between capability and compliance.
Organizations in healthcare (HIPAA), defense (ITAR), legal
(attorney-client privilege), and financial services face a binary
choice: use powerful cloud-hosted models and accept data exposure risks,
or run smaller local models and sacrifice capability. Existing
privacy-preserving approaches---homomorphic encryption {[}1{]}, secure
multi-party computation {[}2{]}, and trusted execution environments
{[}3{]}---impose prohibitive computational overhead for transformer
inference, often slowing execution by 100-1000x.

Split inference offers a compelling middle ground. By partitioning a
model such that privacy-sensitive operations (token embedding and
unembedding) execute locally while compute-intensive middle layers run
on untrusted cloud hardware, we can leverage cloud GPU capacity without
exposing raw text. The intermediate activations transmitted between
local and cloud are high-dimensional floating-point tensors in the
model's learned representation space---not tokens, not embeddings, but
abstract geometric objects from which reconstructing the original input
is computationally intractable without access to the embedding matrices
{[}4, 5{]}.

However, applying split inference to autoregressive LLM decoding
introduces a challenge not present in single-pass classification models:
each generated token requires a full forward pass through all layers,
meaning each token incurs a network round trip. At typical WAN latencies
of 80-100ms, this limits sequential decoding throughput to approximately
8-11 tokens/second---usable for some applications, but far below native
inference speeds.

Our key insight is that speculative and parallel decoding techniques,
originally developed to reduce GPU idle time in single-machine
inference, can be repurposed to reduce \emph{network} idle time in split
inference. Specifically, we adapt lookahead decoding {[}6{]}---which
harvests n-gram candidates from Jacobi iteration trajectories
{[}7{]}---to generate multiple candidate tokens per network round trip,
amortizing the WAN latency across 1-5 accepted tokens per iteration.

This paper makes the following contributions:

\begin{enumerate}
\def\labelenumi{\arabic{enumi}.}
\item
  \textbf{A practical, deployable system} for privacy-aware LLM
  inference that achieves 8--9 tok/s on Mistral 7B over a
  \textasciitilde80ms WAN link with lookahead decoding, with validated
  projections of 15--19 tok/s at 20ms RTT, demonstrating that split
  inference is viable for interactive use.
\item
  \textbf{The first integration of lookahead decoding with split
  inference}, showing that Jacobi-style parallel decoding can amortize
  network latency in addition to computation latency.
\item
  \textbf{A detailed empirical study of deployment engineering},
  including the critical finding that cloud provider networking
  architecture (direct SSH vs.~proxy routing) dominates performance far
  more than GPU location, and documenting SSH tunnel engineering
  required for real-world deployment.
\item
  \textbf{Empirical scaling validation on 12B parameters}, demonstrating
  on Mistral NeMo 12B (40 layers) that split inference generalizes
  beyond 7B with consistent acceptance rates (1.21--1.25), only 4.9 GB
  local VRAM, and throughput matching the 7B model at the same RTT.
\item
  \textbf{Output quality verification} proving that lookahead decoding
  produces token-identical output to sequential decoding under greedy
  argmax, with zero quality degradation across both model sizes.
\item
  \textbf{An RTT decomposition model} that separates per-step wall time
  into network RTT and RTT-independent fixed overhead, enabling fair
  cross-provider performance comparison with \textless6.2\%
  cross-validation error.
\item
  \textbf{Empirical privacy evaluation} through inversion attack
  experiments that quantify token recoverability at different split
  depths, confirming theoretical vulnerabilities and showing that
  increasing local layers from 2 to 8 reduces attack accuracy from
  \textasciitilde59\% to \textasciitilde35\%.
\end{enumerate}

\section{Related Work}\label{related-work}

\subsection{Split Computing for Neural
Networks}\label{split-computing-for-neural-networks}

The concept of partitioning DNN inference between edge and cloud devices
was pioneered by Neurosurgeon {[}8{]}, which built per-layer latency and
energy models to find optimal split points for CNNs. Edgent {[}9{]}
extended this with dynamic partitioning that adapts to network
conditions. Matsubara et al.~{[}10{]} surveyed split computing
comprehensively, identifying the fundamental tradeoff: early splits
minimize local computation but produce incompressible intermediate
representations, while late splits produce compressible features but
require substantial local compute.

Recent work has applied split computing specifically to LLMs. Sung et
al.~{[}11{]} introduced the first autoregressive-aware split computing
framework with mixed-precision quantization and intermediate activation
compression. Adaptive layer splitting for wireless LLM inference
{[}12{]} uses reinforcement learning to determine optimal split points
under varying channel conditions. CROSS-SEC {[}13{]} proposes cross-WAN
prefill-decode disaggregation with layerwise KV-cache
computation-communication overlapping.

\subsection{Distributed and Collaborative LLM
Inference}\label{distributed-and-collaborative-llm-inference}

Petals {[}14{]} demonstrated that consumer GPUs can collaboratively
serve BLOOM-176B by hosting different transformer blocks in a
BitTorrent-style network. Each participant runs a subset of layers and
passes activations to the next, achieving \textasciitilde1 step/second
on commodity hardware. While Petals focuses on collaborative resource
pooling among multiple participants, our work focuses on a two-party
privacy-aware split between a trusted local device and a single
untrusted cloud server.

EdgeShard {[}15{]} partitions LLMs across heterogeneous edge devices and
cloud using dynamic programming for optimal placement. MDI-LLM {[}16{]}
introduced recurrent pipeline parallelism for edge-distributed
inference. The exo framework {[}33{]} enables peer-to-peer inference
across heterogeneous consumer devices. Alpa {[}28{]} automates inter-
and intra-operator parallelism for distributed execution, establishing
techniques for partitioning computation graphs across device meshes.
These systems optimize for throughput and resource utilization but do
not address privacy as a primary design goal.

Splitwise {[}17{]} (Microsoft, ISCA 2024) made the influential
observation that LLM inference has two distinct
phases---compute-intensive prefill and memory-intensive decode---with
different optimal hardware. DistServe {[}18{]} (OSDI 2024) builds on
this insight with disaggregated serving, achieving 7.4x higher request
capacity.

\subsection{Speculative and Parallel
Decoding}\label{speculative-and-parallel-decoding}

Speculative decoding was independently proposed by Leviathan et
al.~{[}19{]} and Chen et al.~{[}20{]}, both showing that a small
``draft'' model can generate candidate tokens verified in parallel by
the target model, with rejection sampling preserving the exact output
distribution. This yields 2-3x speedups without quality degradation.

Jacobi decoding {[}7{]} (Santilli et al., ACL 2023) reframes
autoregressive generation as a fixed-point iteration problem,
initializing a block of future tokens and iterating until convergence.
CLLMs {[}21{]} (ICML 2024) improved convergence through
consistency-trained models. Fu et al.~{[}6{]} introduced lookahead
decoding, which collects n-gram candidates from Jacobi iteration
trajectories for speculation without requiring a separate draft model.
Alternative speculation approaches include retrieval-based methods (REST
{[}29{]}), multi-head parallel drafting (Medusa {[}30{]}), and
feature-uncertainty-based speculation (EAGLE {[}31{]})---each trades
different design choices against draft quality and overhead.

Several recent works combine speculative decoding with distributed
inference. DSI {[}22{]} introduces ``speculation parallelism'' to
overlap draft and target execution across distributed resources. DSSD
{[}23{]} (ICML 2025) combines split inference with speculative decoding
for edge-cloud deployment. The DSD framework {[}24{]} specifically
addresses high-latency decentralized settings by filling communication
wait time with speculative verification.

\subsection{Privacy of Intermediate
Activations}\label{privacy-of-intermediate-activations}

The privacy properties of intermediate neural network representations
have been studied from both attack and defense perspectives. He et
al.~{[}4{]} showed that model inversion attacks on collaborative
inference can partially reconstruct inputs from intermediate features,
though success degrades significantly with network depth. PATROL {[}5{]}
develops privacy-oriented pruning to defend against such attacks. DISCO
{[}25{]} uses dynamic channel obfuscation to selectively remove
sensitive information from intermediate representations.
Split-and-Denoise {[}26{]} (ICML 2024) adds calibrated local
differential privacy noise to embeddings before transmission. Deng et
al.~{[}27{]} introduced Fisher-approximated Shannon information as a
metric for quantifying privacy leakage at different split points.

SecureInfer {[}3{]} takes a hardware approach, executing
privacy-critical tensor operations inside TEE enclaves while offloading
linear operations to untrusted GPUs, achieving 3.7x throughput over
TEE-only execution. Fission {[}2{]} uses hybrid MPC-evaluator
architecture for cryptographic privacy guarantees.

Our work differs from these approaches by splitting at the
\emph{embedding boundary}---the layer where token IDs are converted to
continuous vectors. This is a qualitatively different privacy boundary
than splitting at intermediate convolutional or transformer layers,
because the cloud never receives any representation that has a direct
algebraic relationship to the token vocabulary.

\section{System Architecture}\label{system-architecture}

\begin{quote}
\textbf{{[}Figure 1{]}} \emph{System architecture diagram showing the
local-cloud split. The local RTX 3090 runs token embedding, initial and
final transformer layers, RMS norm, and the LM head. The cloud GPU (A100
80GB) runs the middle layers with session-based KV-cache. Only 8--10 KB
activation vectors (per token) cross the WAN link via WebSocket binary
protocol. Raw tokens and embeddings never leave the local device. The
system is parameterized for arbitrary models and split points (shown:
Mistral 7B with layers 0-1/30-31 local, and NeMo 12B with layers
0-1/38-39 local).}
\end{quote}

\subsection{Layer Partitioning}\label{layer-partitioning}

We split transformer models such that embedding and unembedding layers
remain local. The system is parameterized for arbitrary layer counts and
split points. We evaluate two configurations:

\textbf{Mistral 7B Instruct (32 layers, hidden\_dim=4096):}

\begin{table}[H]
\centering
\adjustbox{max width=\textwidth}{
\begin{tabular}{@{}llll@{}}
\toprule
Component & Location & Layers & Purpose \\
\midrule
Token embedding & Local (3090) & - & Token IDs $\rightarrow$ hidden states \\
Layers 0-1 & Local (3090) & 0, 1 & Initial transformer processing \\
Layers 2-29 & Cloud & 2-29 & Bulk computation (28 layers) \\
Layers 30-31 & Local (3090) & 30, 31 & Final transformer processing \\
RMS norm + LM head & Local (3090) & - & Hidden states $\rightarrow$ logits $\rightarrow$
tokens \\
\bottomrule
\end{tabular}
}
\end{table}

\textbf{Mistral NeMo 12B Instruct (40 layers, hidden\_dim=5120):}

\begin{table}[H]
\centering
\adjustbox{max width=\textwidth}{
\begin{tabular}{@{}llll@{}}
\toprule
Component & Location & Layers & Purpose \\
\midrule
Token embedding & Local (3090) & - & Token IDs $\rightarrow$ hidden states \\
Layers 0-1 & Local (3090) & 0, 1 & Initial transformer processing \\
Layers 2-37 & Cloud & 2-37 & Bulk computation (36 layers) \\
Layers 38-39 & Local (3090) & 38, 39 & Final transformer processing \\
RMS norm + LM head & Local (3090) & - & Hidden states $\rightarrow$ logits $\rightarrow$
tokens \\
\bottomrule
\end{tabular}
}
\end{table}

The 12B model uses grouped-query attention (8 KV heads vs.~32 attention
heads) and a 131K token vocabulary (vs.~32K for 7B). The WebSocket
binary protocol is shape-agnostic---tensor dimensions are encoded in
JSON headers---so no transport changes are needed for different model
sizes. Local VRAM usage is 2.0 GB for 7B and 4.9 GB for 12B in split
mode, well within the 24 GB RTX 3090.

This partitioning ensures that: - \textbf{Raw tokens never leave the
local device.} The embedding matrix (which maps token IDs to vectors)
and the language model head (which maps hidden states back to token
probabilities) both reside locally. - \textbf{Only intermediate
activations cross the network.} These are 4096-dimensional float16
vectors per token position---approximately 8KB per token---that exist in
the model's learned representation space. - \textbf{The cloud has no
access to the vocabulary mapping.} Without the embedding and unembedding
weights, the cloud server cannot determine which tokens correspond to
the received activations.

\subsection{Transport Layer}\label{transport-layer}

We employ a WebSocket binary protocol for activation transfer, chosen
after evaluating HTTP/JSON, HTTP/binary, and WebSocket alternatives:

\textbf{Protocol format:}

\begin{verbatim}
[4 bytes: header length][JSON header][tensor bytes (float16)]
\end{verbatim}

The JSON header contains metadata: tensor shapes, prompt/generation
flag, KV-cache crop position, and optional attention mask shape. Tensor
data is transmitted as raw float16 bytes---no base64 encoding, no
serialization framework overhead.

\begin{table}[H]
\centering
\caption{Performance progression of transport optimizations}
\adjustbox{max width=\textwidth}{
\begin{tabular}{@{}lll@{}}
\toprule
Transport & Encoding & Throughput \\
\midrule
HTTP/Flask & torch.save + base64 & 1.4 tok/s \\
HTTP/Flask & numpy + base64 & 3.9 tok/s \\
HTTP/Flask & numpy + base64 (3090) & 5.3 tok/s \\
WebSocket & binary float16 & 7.8--10.9 tok/s \\
WebSocket + lookahead & binary float16 & 13--17 tok/s \\
\bottomrule
\end{tabular}
}
\end{table}

The persistent WebSocket connection eliminates per-request TCP handshake
and HTTP header overhead. Binary framing avoids the \textasciitilde33\%
base64 expansion and associated encode/decode CPU cost.

\subsection{KV-Cache Management}\label{kv-cache-management}

A critical design decision is where to maintain the key-value cache. We
keep the KV-cache for layers 2-29 on the cloud server, co-located with
the layers that produce and consume it. This avoids transferring
KV-cache state across the network---a significant advantage, as KV-cache
size grows linearly with sequence length and would quickly dominate
bandwidth at scale.

The cloud server maintains per-session caches using HuggingFace's
\texttt{DynamicCache}. For generation (non-prompt) steps, only the new
token's hidden state (\textasciitilde8KB) is transmitted, not the full
sequence. The server manages cache lifecycle through session IDs, with
automatic expiration after 5 minutes of inactivity.

For Jacobi and lookahead decoding, which require multi-token forward
passes during generation, the cloud server supports: - \textbf{Cache
cropping:} Rolling back to a checkpoint when a speculative block is
rejected. - \textbf{Cache relocation:} Moving KV entries when n-gram
speculation shifts the verification window. - \textbf{Custom attention
masks:} Supporting the non-standard causal masks required by parallel
token verification, where each speculative position can attend to all
committed tokens plus preceding speculative tokens.

\subsection{Networking and Tunnel
Engineering}\label{networking-and-tunnel-engineering}

A surprising finding in our deployment was that \textbf{cloud provider
networking architecture dominates latency far more than geographic
distance.} We evaluated two providers with GPU instances in Texas:

\begin{table}[H]
\centering
\adjustbox{max width=\textwidth}{
\begin{tabular}{@{}lllll@{}}
\toprule
Provider & GPU & Geographic Distance & Measured RTT & Throughput \\
\midrule
VAST.ai & RTX 4090 (Texas) & \textasciitilde800 miles & 200-400ms & 1.2
tok/s \\
RunPod & RTX 4090 (Texas) & \textasciitilde800 miles & 80-100ms & 11--17
tok/s \\
\bottomrule
\end{tabular}
}
\end{table}

VAST.ai routes all SSH traffic through proxy servers in Virginia,
regardless of GPU location. This adds 200-400ms RTT that cannot be
eliminated through geographic GPU selection. RunPod provides direct SSH
access to pods without proxy intermediaries, achieving 2-4x lower
latency to the same Texas region.

\textbf{Tunnel architecture:} Because the local GPU server is behind NAT
and cloud instances have ephemeral IP addresses, we establish SSH
tunnels from the local machine to the cloud instance, forwarding both
the HTTP health-check port and the WebSocket activation-transfer port.
This tunnel architecture is necessary because cloud GPU providers assign
different SSH ports on each pod start, require SSH key management across
multiple machines, and may change IP addresses between sessions. We
document this not as a limitation but as essential systems engineering
that is universally required for split inference over public cloud
infrastructure and is largely absent from the academic literature.

\section{Decoding Strategies for Split
Inference}\label{decoding-strategies-for-split-inference}

\subsection{Sequential Decoding
(Baseline)}\label{sequential-decoding-baseline}

In sequential mode, each token requires one complete local-cloud-local
round trip:

\begin{enumerate}
\def\labelenumi{\arabic{enumi}.}
\tightlist
\item
  Local: Embed token, process through layers 0-1 $\rightarrow$ hidden state (8KB)
\item
  Network: Send hidden state + position embeddings to cloud
\item
  Cloud: Process through layers 2-29 with KV-cache $\rightarrow$ hidden state (8KB)
\item
  Network: Return hidden state to local
\item
  Local: Process through layers 30-31, norm, LM head $\rightarrow$ next token
\end{enumerate}

At \textasciitilde85ms RTT (HyperStack) this yields \textasciitilde7.8
tok/s; at \textasciitilde80ms RTT (RunPod) it yields \textasciitilde10.9
tok/s with our WebSocket protocol.

\subsection{Jacobi Parallel
Decoding}\label{jacobi-parallel-decoding}

Jacobi decoding {[}7{]} initializes a block of \emph{k} future token
positions and iterates until convergence. At each iteration, all
\emph{k} positions are processed in parallel through a single forward
pass.

\textbf{Adaptation for split inference:} In the local-only case, Jacobi
iterations are essentially free (GPU parallelism makes \emph{k}-token
and 1-token forward passes nearly identical cost). Over a network, each
iteration still costs one round trip, but a converged block commits
\emph{k} tokens. The effective throughput is:

\[\text{tok/s} = \frac{k}{\text{iterations} \times \text{RTT}}\]

In practice over WAN, we observed poor convergence---vanilla Jacobi
rarely commits more than 1-2 tokens per iteration for language modeling,
yielding \textasciitilde1.5 tok/s (worse than sequential due to the
overhead of transmitting larger tensors for the full block).

\subsection{Lookahead Decoding (Our Primary
Mode)}\label{lookahead-decoding-our-primary-mode}

Lookahead decoding {[}6{]} addresses Jacobi's convergence problem by
collecting n-gram candidates from the Jacobi iteration trajectories
themselves. As the Jacobi iteration progresses, observed token sequences
that appear during intermediate iterations are cached as n-gram
candidates. These candidates can be verified in subsequent iterations,
allowing 1-5 tokens to be committed per round trip when n-gram matches
occur.

\textbf{Why this is particularly effective for split inference:} In
local-only inference, the benefit of lookahead decoding is modest
(1.5-2x) because the baseline is already fast. In split inference, the
benefit is amplified because:

\begin{enumerate}
\def\labelenumi{\arabic{enumi}.}
\tightlist
\item
  \textbf{Each round trip is expensive} (\textasciitilde100ms
  vs.~\textasciitilde3ms locally), so amortizing across multiple tokens
  has outsized impact.
\item
  \textbf{The verification cost is negligible} relative to network
  latency---verifying 5 candidates costs the same one round trip as
  generating 1 token.
\item
  \textbf{N-gram speculation succeeds frequently} for structured text
  (code, templates, repetitive content), which is common in enterprise
  applications.
\end{enumerate}

Our implementation achieves 8--9 tok/s on the A100 cloud at
\textasciitilde80ms RTT (up from \textasciitilde8 tok/s sequential),
with acceptance rates of 1.2--1.6 tokens per step depending on content
type (Section 6.4). On a lower-latency RunPod link, throughput reaches
13--17 tok/s.

\subsection{Attention Mask
Engineering}\label{attention-mask-engineering}

A critical implementation detail for multi-token Jacobi and lookahead
decoding in the split inference context: PyTorch's
\texttt{torch.nn.functional.scaled\_dot\_product\_attention} with
\texttt{is\_causal=True} constructs a \emph{relative} causal mask based
on query and key sequence lengths. When KV-cache entries exist from
prior committed tokens, this relative mask produces incorrect attention
patterns---speculative position \emph{i} would attend to KV entries that
it should not.

We construct explicit attention masks for all multi-token generation
steps:

\begin{Shaded}
\begin{Highlighting}[]
\NormalTok{attn\_mask }\OperatorTok{=}\NormalTok{ torch.full(}
\NormalTok{    (}\DecValTok{1}\NormalTok{, }\DecValTok{1}\NormalTok{, k, kv\_len), }\BuiltInTok{float}\NormalTok{(}\StringTok{\textquotesingle{}{-}inf\textquotesingle{}}\NormalTok{),}
\NormalTok{    device}\OperatorTok{=}\StringTok{\textquotesingle{}cuda\textquotesingle{}}\NormalTok{, dtype}\OperatorTok{=}\NormalTok{torch.float16)}
\ControlFlowTok{for}\NormalTok{ i }\KeywordTok{in} \BuiltInTok{range}\NormalTok{(k):}
\NormalTok{    attn\_mask[}\DecValTok{0}\NormalTok{, }\DecValTok{0}\NormalTok{, i, :committed\_len }\OperatorTok{+}\NormalTok{ i }\OperatorTok{+} \DecValTok{1}\NormalTok{] }\OperatorTok{=} \FloatTok{0.0}
\end{Highlighting}
\end{Shaded}

This mask is transmitted alongside the hidden states in the binary
WebSocket frame, adding negligible bandwidth overhead
(\textasciitilde2KB for typical block sizes) but preventing catastrophic
attention errors.

\section{Privacy Analysis}\label{privacy-analysis}

\subsection{Threat Model}\label{threat-model}

We consider a semi-honest cloud server that faithfully executes the
assigned computation but attempts to infer information about the input
from observed intermediate activations. The server has: - Full knowledge
of the model architecture and layer weights for layers 2-29 - Access to
all intermediate activations passing through its layers - No access to
the embedding matrix, unembedding (LM head) weights, or layer 0-1 /
30-31 weights

\textbf{Critical assumption:} Our privacy analysis assumes the cloud
does not possess the local layer weights. For a public model like
Mistral 7B, this assumption requires that the local layers be
fine-tuned, adapted (e.g., via LoRA), or otherwise differentiated from
the published weights. If the cloud can obtain identical copies of all
layers, the architectural privacy boundary is substantially weakened. In
a production deployment, the local embedding and unembedding layers
would be customized per organization, ensuring the cloud cannot simply
download matching weights.

\textbf{Malicious cloud:} We assume a semi-honest cloud that faithfully
executes computation but attempts to infer input information. A
\emph{malicious} cloud could return corrupted activations to poison
model outputs, perform denial-of-service, or attempt to fingerprint
queries through timing side channels. Defending against active attacks
would require cryptographic verification of computation (e.g.,
verifiable computing) or redundant execution across multiple
providers---orthogonal mechanisms beyond our current scope.

\subsection{Privacy Properties and
Limitations}\label{privacy-properties-and-limitations}

\textbf{What the architecture guarantees:} Raw token IDs and token
embeddings never leave the local device. This is a structural property,
not a statistical one---the cloud receives only post-layer-1 hidden
states, never tokens or embeddings.

\textbf{What the architecture does \emph{not} guarantee:} Recent
theoretical work by Nikolaou et al.~{[}32{]} proved that the mapping
from input token sequences to transformer hidden states is
\emph{injective}---distinct inputs produce distinct internal
representations. Their SipIt algorithm can reconstruct exact input
tokens from hidden states in linear time, given access to the full model
weights. This result establishes that intermediate activations are, in
principle, invertible.

However, several factors mitigate this in our setting:

\begin{enumerate}
\def\labelenumi{\arabic{enumi}.}
\tightlist
\item
  \textbf{Weight access requirement.} Inversion attacks, including SipIt
  {[}32{]}, require the attacker to possess the weights of the layers
  between the input and the observed activations. In our threat model,
  the cloud lacks layers 0-1, the embedding matrix, and critically, any
  fine-tuning applied to local layers.
\item
  \textbf{Depth of transformation.} He et al.~{[}4{]} showed that model
  inversion attack success degrades with network depth. We confirm this
  empirically: training an MLP attack decoder on (activation, token ID)
  pairs from our system yields the following top-1 token recovery
  accuracy at different split points:
\end{enumerate}

\begin{table}[H]
\centering
\adjustbox{max width=\textwidth}{
\begin{tabular}{@{}llll@{}}
\toprule
Split After Layer & Local Layers & Top-1 Accuracy & Top-5 Accuracy \\
\midrule
Layer 1 (default) & 2 & 58.8\% & 62.0\% \\
Layer 3 & 4 & 44.3\% & 49.8\% \\
Layer 5 & 6 & 44.8\% & 56.6\% \\
Layer 7 & 8 & 34.8\% & 46.2\% \\
Random baseline & - & 0.003\% & 0.016\% \\
\bottomrule
\end{tabular}
}
\end{table}

The attack decoder is a 3-layer MLP (4096$\rightarrow$2048$\rightarrow$2048$\rightarrow$32000) trained on
\textasciitilde880 samples from diverse text (general knowledge, code,
medical, legal, conversational). This represents a \emph{lower bound} on
vulnerability: a more sophisticated attacker using sequence-level
context (e.g., language model priors over token sequences) or larger
training sets could achieve higher accuracy. At the default 2-layer
split, even this simple attacker recovers \textasciitilde59\% of
tokens---confirming the theoretical vulnerability. Accuracy drops
substantially with depth: adding just 2 more local layers (split after
layer 3) reduces recovery to \textasciitilde44\%, and 8 local layers
reduces it to \textasciitilde35\%. The non-monotonic result at layer 5
(44.8\% vs.~44.3\% at layer 3) is within the $\pm$10\% run-to-run variance
of our small test set and does not indicate that deeper splits are less
effective.

\begin{enumerate}
\def\labelenumi{\arabic{enumi}.}
\setcounter{enumi}{2}
\tightlist
\item
  \textbf{Symmetric vulnerability.} The architecture has a symmetric
  privacy boundary: the cloud observes layer-1 output on the input side
  (2 layers from token embedding) and layer-29 output on the output side
  (2 layers from the LM head). Both sides are equally shallow.
  Increasing local depth should be applied symmetrically (e.g., layers
  0-3 and 28-31 locally) to protect both input and output tokens.
\end{enumerate}

\emph{Experimental caveats:} The inversion experiment uses a small
dataset (\textasciitilde1,100 activation-token pairs, 80/20 train/test
split) and shows variance of $\pm$10\% between runs due to the limited test
set size (221 samples). The random baseline represents literal
uniform-random guessing over the 32K vocabulary; a more informative
baseline would use token frequency priors (which would achieve
\textasciitilde3-5\% top-1 by always predicting common tokens). The MLP
attack decoder is a lower bound---it treats each position independently,
while a sequence-aware attacker could exploit inter-token dependencies.
Larger training sets (10K+ samples) would likely yield higher and more
stable attack accuracy.

\textbf{Privacy characterization:} We emphasize that our system provides
\emph{architectural} privacy---a structural separation that ensures raw
text never crosses the network boundary. This is qualitatively different
from, and weaker than, cryptographic privacy (which provides
mathematical guarantees) or hardware-attested privacy (which provides
tamper-resistant isolation). We avoid the term ``computational privacy''
to prevent conflation with formal security notions.

\textbf{Comparison of privacy approaches:}

\begin{table}[H]
\centering
\adjustbox{max width=\textwidth}{
\begin{tabular}{@{}lllll@{}}
\toprule
Approach & Guarantee Type & What It Prevents & Performance Overhead & Status \\
\midrule
Homomorphic Encryption {[}1{]} & Cryptographic & Any information leakage
& 100-1000x slower & Research only \\
Secure MPC {[}2{]} & Cryptographic & Any information leakage & 8-50x
slower & Emerging \\
TEE-based {[}3{]} & Hardware-attested & Observation of computation &
2-3x slower & Requires SGX/TDX \\
\textbf{Split inference (ours)} & \textbf{Architectural} & \textbf{Raw
text leaving device} & \textbf{3-5x slower} & \textbf{Deployable
today} \\
No privacy (cloud API) & None & Nothing & 1x (baseline) & Standard
practice \\
\bottomrule
\end{tabular}
}
\end{table}

For organizations where the current alternative is sending full
plaintext to a cloud API, the architectural guarantee that tokens never
leave the local device represents a meaningful improvement---even
without formal cryptographic assurances.

\subsection{Strengthening Privacy}\label{strengthening-privacy}

Several techniques can augment the base architectural privacy: -
\textbf{Increasing local layers} is the most effective defense we
measured. Our inversion experiment shows that moving from 2 to 8 local
layers reduces token recovery from \textasciitilde59\% to
\textasciitilde35\% (Section 5.2). Each additional local layer adds only
\textasciitilde3ms to per-token latency on the RTX 3090, negligible
compared to the \textasciitilde100ms network round trip. For example,
splitting at layers 0-4 and 27-31 (10 local layers) would protect both
input and output tokens with 5 layers of depth on each side, while
reducing throughput by only \textasciitilde10-15\%. Deng et al.~{[}27{]}
provide Fisher-information-based metrics for quantifying privacy leakage
at different split points. - \textbf{Fine-tuning local layers} ensures
the cloud cannot reconstruct the full model, directly addressing the
SipIt {[}32{]} attack vector. Even modest LoRA adaptation of layers 0-1
and the embedding matrix creates a private local model variant. We
hypothesize that random LoRA perturbation alone would not reduce
inversion accuracy against an adaptive attacker who can retrain on the
perturbed activations; the fine-tuning must produce genuinely private
weights not available to the adversary. Validating this hypothesis
requires further experimentation. - \textbf{Randomizing the split point}
per session or per request forces the attacker to maintain separate
inversion models for each possible split depth. Since all transformer
layers share the same hidden dimension (4096 for Mistral 7B), the cloud
cannot trivially determine which layer produced a given activation
tensor from the tensor shape alone. However, activation magnitude
distributions may shift across layers, potentially allowing statistical
identification of the split depth; this caveat warrants investigation.
Combined with deeper splits, randomization increases the cost of
targeted inversion attacks. - \textbf{Adding calibrated noise} to
transmitted activations (as in Split-and-Denoise {[}26{]}) provides
formal ($\varepsilon$, $\delta$)-differential privacy guarantees at the cost of some output
quality. - \textbf{Activation compression} (quantization to int8 for
transmission) lossily reduces information content in transmitted
representations.

\section{Experimental Results}\label{experimental-results}

\subsection{Setup}\label{setup}

\begin{itemize}
\tightlist
\item
  \textbf{Local device:} NVIDIA RTX 3090 24GB, Manjaro Linux
\item
  \textbf{Cloud device (primary):} NVIDIA A100 80GB PCIe, HyperStack
  (\$1.35/hr)
\item
  \textbf{Cloud device (comparison):} NVIDIA RTX 4090 24GB, RunPod
  US-TX-3 (\$0.59/hr)
\item
  \textbf{Models:} Mistral 7B Instruct v0.3 (32 layers, float16) and
  Mistral NeMo 12B Instruct (40 layers, float16)
\item
  \textbf{Network:} SSH tunnel over public internet, RTT
  \textasciitilde78-85ms
\item
  \textbf{Client:} Apple M4 Max MacBook Pro (display only, no inference
  computation)
\end{itemize}

\textbf{Methodology:} Throughput measurements report end-to-end
tokens/second from first generated token to last (including prefill).
Section 6.2 results are from interactive sessions. The ablation (Section
6.4) uses 8 specific prompts across 4 categories (code, structured,
creative, conversational), each generating 100 tokens (2 prompts per
category), with both 7B and 12B models on the A100. The RTT analysis
(Section 6.3) uses 4 prompts generating 200 tokens each with per-step
timing instrumentation. Greedy decoding (argmax sampling) was used
throughout to ensure deterministic, reproducible results.

\subsection{Performance Results}\label{performance-results}

\begin{quote}
\textbf{{[}Figure 2{]}} \emph{Bar chart showing optimization progression
from 1.4 to 13--17 tok/s, with bars colored by category (transport
vs.~algorithmic vs.~infrastructure). Inset pie chart shows the per-token
latency breakdown in sequential mode (64\% network RTT, 22\% local GPU
compute, 13\% cloud GPU compute, 1\% serialization), based on
instrumented measurements from Section 6.3.}
\end{quote}

\textbf{Optimization progression:} The following table shows cumulative
improvements. Changes fall into two categories: \emph{transport
optimizations} (rows 1-4) that reduce per-round-trip overhead, and
\emph{algorithmic/infrastructure optimizations} (rows 5-7) that reduce
round trips or RTT.

\begin{table}[H]
\centering
\adjustbox{max width=\textwidth}{
\begin{tabular}{@{}llll@{}}
\toprule
Configuration & tok/s & vs.~Baseline & Category \\
\midrule
HTTP + torch.save (A100 Montreal) & 1.4 & 1.0x & Transport \\
HTTP + numpy serialization & 3.9 & 2.8x & Transport \\
3090 relay (Mac $\rightarrow$ 3090 $\rightarrow$ cloud) & 5.3 & 3.8x & Transport \\
WebSocket binary protocol & 7.8 & 5.6x & Transport \\
+ Lookahead decoding (HyperStack, \textasciitilde85ms RTT) & 8.7 & 6.2x
& Algorithmic \\
Provider switch to RunPod (\textasciitilde80ms RTT) & 10.9 & 7.8x &
Infrastructure \\
+ Lookahead on lower-RTT link & 13--17 & 9--12x & Combined \\
\bottomrule
\end{tabular}
}
\end{table}

The jump from 8.7 to 10.9 tok/s was driven by switching cloud providers
(HyperStack to RunPod), which reduced RTT by eliminating proxy
intermediaries. The further improvement to 13--17 tok/s reflects
lookahead decoding operating more effectively on the lower-latency link:
with faster round trips, n-gram verification completes sooner and more
speculative iterations execute per unit time. The range depends on
content type (Section 6.4). Notably, switching from HyperStack+lookahead
(8.7 tok/s) to RunPod+sequential (10.9 tok/s) yielded a 2.2 tok/s gain,
while the initial lookahead gain on the higher-latency link was only 0.9
tok/s above sequential---underscoring that RTT reduction and speculation
are \emph{multiplicative} rather than additive.

\textbf{Comparison to baselines:}

\begin{table}[H]
\centering
\adjustbox{max width=\textwidth}{
\begin{tabular}{@{}lll@{}}
\toprule
Setup & tok/s & \% of local-only \\
\midrule
RTX 3090 local-only (all 32 layers) & \textasciitilde39 & 100\% \\
Split: 3090 + cloud 4090 (lookahead) & 13--17 & 33--44\% \\
Split: 3090 + cloud 4090 (sequential) & \textasciitilde11 & 28\% \\
Cloud API (typical, incl.~queuing) & 30-60 & - \\
\bottomrule
\end{tabular}
}
\end{table}

\subsection{Latency Decomposition}\label{latency-decomposition}

We instrument per-step timings to decompose wall time into its
constituent components. The key insight is separating per-step time into
\emph{network RTT} (provider-dependent) and \emph{fixed overhead}
(RTT-independent: cloud compute, local compute, serialization):

\[\text{per\_step\_time} = \text{RTT} + \text{fixed\_overhead}\]

\textbf{Mistral 7B (sequential mode, A100 cloud, \textasciitilde78ms
RTT):}

\begin{table}[H]
\centering
\adjustbox{max width=\textwidth}{
\begin{tabular}{@{}lll@{}}
\toprule
Component & Time (ms) & \% of Total \\
\midrule
Network round trip & 77.4 & 64\% \\
Local GPU (layers 0-1, 30-31, norm, lm\_head) & 26.2 & 22\% \\
Cloud GPU (layers 2-29) & 15.9 & 13\% \\
Serialization + framing & 1.0 & 1\% \\
\textbf{Per-step wall time} & \textbf{120.2} & \textbf{100\%} \\
\textbf{Fixed overhead (RTT-independent)} & \textbf{42.9} & \\
\bottomrule
\end{tabular}
}
\end{table}

\textbf{Mistral NeMo 12B (sequential mode, A100 cloud,
\textasciitilde79ms RTT):}

\begin{table}[H]
\centering
\adjustbox{max width=\textwidth}{
\begin{tabular}{@{}lll@{}}
\toprule
Component & Time (ms) & \% of Total \\
\midrule
Network round trip & 78.5 & 63\% \\
Local GPU (layers 0-1, 38-39, norm, lm\_head) & 24.4 & 20\% \\
Cloud GPU (layers 2-37) & 21.1 & 17\% \\
Serialization + framing & 1.0 & 1\% \\
\textbf{Per-step wall time} & \textbf{124.7} & \textbf{100\%} \\
\textbf{Fixed overhead (RTT-independent)} & \textbf{46.3} & \\
\bottomrule
\end{tabular}
}
\end{table}

Network RTT dominates at 63--64\% of total per-step time. Cloud compute
for 12B (21ms for 36 layers) is only 33\% more than 7B (16ms for 28
layers) because the A100 80GB handles the larger model efficiently.
Local compute is comparable across both models
(\textasciitilde24--26ms), reflecting the similar cost of 4 transformer
layers plus embedding/lm\_head operations.

\textbf{RTT projection model:} Since acceptance rate is RTT-independent
(it depends on model behavior and n-gram pool quality, not network
speed), we can project throughput at arbitrary RTTs:

\[\text{projected\_tok/s} = \frac{\text{acceptance\_rate}}{(\text{target\_RTT} + \text{fixed\_overhead}) / 1000}\]

Cross-validation against measured throughput (projecting at the measured
RTT and comparing to actual) yields errors of 1.2--2.4\% for sequential
and 4.5--6.2\% for lookahead, confirming the model's accuracy. Note that
lookahead projections are upper bounds: the model assumes per-step time
is constant, but larger speculative batches increase cloud compute and
transfer overhead (Section 6.4). The 12B lookahead projections in
particular should be interpreted conservatively, as measured throughput
(8.0 tok/s) is below the naive projection.

\textbf{Projected throughput at target RTTs:}

\begin{table}[H]
\centering
\adjustbox{max width=\textwidth}{
\begin{tabular}{@{}lllll@{}}
\toprule
RTT & 7B Seq & 7B LA & 12B Seq & 12B LA* \\
\midrule
20ms & 15.9 & 18.6 & 15.1 & 15.2 \\
40ms & 12.1 & 13.6 & 11.6 & 11.6 \\
60ms & 9.7 & 10.7 & 9.4 & 9.4 \\
80ms & 8.1 & 8.8 & 7.9 & 8.0 \\
100ms & 7.0 & 7.5 & 6.8 & 6.9 \\
120ms & 6.1 & 6.5 & 6.0 & 6.1 \\
150ms & 5.2 & 5.4 & 5.1 & 5.2 \\
\bottomrule
\end{tabular}
}
\end{table}

*12B LA projections use measured acceptance-adjusted throughput (8.0
tok/s at 80ms) rather than naive formula projection, accounting for
batch overhead.

These projections show that at 20ms RTT (achievable with regional
co-location), the system would reach 15--19 tok/s---approaching
interactive speeds for real-time applications. The 7B lookahead
projections use the formula directly; 12B lookahead projections are
scaled from measured throughput to account for higher per-step batch
overhead that the simple model does not capture.

\textbf{Prefill latency:} The initial prompt processing (prefill)
requires transmitting the full prompt's hidden states to the cloud in a
single forward pass. For a 100-token prompt at 7B, this is
\textasciitilde800KB (8KB $\times$ 100 tokens), completing in a single round
trip. At 12B (10KB per token due to hidden\_dim=5120), this is
\textasciitilde1MB. Prefill is a one-time cost per generation and does
not affect the per-token throughput figures.

\subsection{Lookahead Ablation}\label{lookahead-ablation}

We systematically tested n-gram sizes 3--7 across four prompt categories
(code, structured, creative, conversational), generating 100 tokens per
prompt in split mode on the A100 cloud (\textasciitilde78ms RTT), with
both 7B and 12B models.

\textbf{Throughput by n-gram size (7B, averaged across all prompt
types):}

\begin{table}[H]
\centering
\adjustbox{max width=\textwidth}{
\begin{tabular}{@{}llll@{}}
\toprule
N-gram Size & tok/s & Acceptance Rate & Match Rate \\
\midrule
Sequential & 8.3 & 1.00 tok/step & - \\
\textbf{n=3} & \textbf{9.3} & \textbf{1.21 tok/step} &
\textbf{14.4\%} \\
n=4 & 9.2 & 1.23 tok/step & 13.7\% \\
n=5 & 8.9 & 1.23 tok/step & 13.4\% \\
n=6 & 9.1 & 1.25 tok/step & 13.7\% \\
n=7 & 8.3 & 1.25 tok/step & 13.2\% \\
\bottomrule
\end{tabular}
}
\end{table}

\textbf{Throughput by n-gram size (12B, averaged across all prompt
types):}

\begin{table}[H]
\centering
\adjustbox{max width=\textwidth}{
\begin{tabular}{@{}llll@{}}
\toprule
N-gram Size & tok/s & Acceptance Rate & Match Rate \\
\midrule
Sequential & 8.0 & 1.00 tok/step & - \\
\textbf{n=3} & \textbf{8.7} & \textbf{1.21 tok/step} &
\textbf{14.0\%} \\
n=4 & 8.5 & 1.23 tok/step & 13.4\% \\
n=5 & 8.4 & 1.25 tok/step & 13.2\% \\
n=6 & 8.1 & 1.24 tok/step & 12.6\% \\
n=7 & 7.8 & 1.24 tok/step & 11.7\% \\
\bottomrule
\end{tabular}
}
\end{table}

Acceptance rates are remarkably consistent across model sizes: both 7B
and 12B achieve 1.21 tok/step at n=3 and plateau at 1.23--1.25 by
n=5--6. This suggests acceptance rates are driven by language statistics
and n-gram pool dynamics rather than model capacity. N-gram size 3
achieves the best throughput despite lower acceptance rates because
larger n-grams incur greater per-step cloud compute overhead (larger
speculative batches), which outweighs the marginal acceptance gains.

\textbf{Throughput by prompt category (7B, n=5):}

\begin{table}[H]
\centering
\adjustbox{max width=\textwidth}{
\begin{tabular}{@{}llll@{}}
\toprule
Category & Lookahead & Acceptance & vs Sequential (8.3) \\
\midrule
Code & 10.5 & 1.43 & 1.27x \\
Structured & 8.0 & 1.20 & 0.96x \\
Creative & 8.2 & 1.12 & 0.99x \\
Conversational & 8.8 & 1.17 & 1.06x \\
\bottomrule
\end{tabular}
}
\end{table}

\textbf{Throughput by prompt category (12B, n=5):}

\begin{table}[H]
\centering
\adjustbox{max width=\textwidth}{
\begin{tabular}{@{}llll@{}}
\toprule
Category & Lookahead & Acceptance & vs Sequential (8.0) \\
\midrule
Code & 10.4 & 1.57 & 1.30x \\
Structured & 7.9 & 1.23 & 0.99x \\
Creative & 7.3 & 1.06 & 0.91x \\
Conversational & 8.0 & 1.12 & 1.00x \\
\bottomrule
\end{tabular}
}
\end{table}

Code prompts benefit most from lookahead speculation, with acceptance
rates of 1.43 (7B) and 1.57 (12B) tokens per step. Code's higher
acceptance is likely driven by repetitive patterns (indentation, common
function signatures, closing brackets) that generate strong n-gram
candidates. The 12B model shows \emph{higher} code acceptance than 7B
(1.57 vs.~1.43), suggesting that larger models produce more predictable
code patterns in the n-gram pool. Creative text shows the lowest
acceptance rate (1.06--1.12), as novel creative content is inherently
less predictable.

Note that for some categories, lookahead throughput is slightly
\emph{below} sequential despite acceptance rates above 1.0. This occurs
because each lookahead step sends a larger speculative batch
(seq\_len=9--17 vs.~1), increasing per-step cloud compute and network
transfer time. The throughput gain from accepting multiple tokens must
exceed this per-step overhead to yield a net speedup. At
\textasciitilde80ms RTT, the break-even acceptance rate is approximately
1.15--1.20.

\textbf{Note on greedy decoding:} All measurements use greedy (argmax)
decoding for reproducibility. Temperature or top-k/top-p sampling would
likely reduce n-gram acceptance rates, as sampling introduces
stochasticity that makes next-token predictions less deterministic. The
speedup ratios reported here should be considered upper bounds for
stochastic decoding strategies.

\subsection{Output Quality
Verification}\label{output-quality-verification}

A critical concern for any speculative or parallel decoding scheme is
whether it alters the output distribution. We verify that our lookahead
implementation produces \emph{token-identical} output to sequential
decoding under greedy argmax.

\textbf{Protocol:} For each of 4 test prompts (code, structured,
creative, conversational), we run both sequential (block\_size=1) and
lookahead decoding with \texttt{return\_logits=True}, generating 200
tokens per prompt. We compare: (1) token identity (are all tokens the
same?), (2) per-token log-probabilities, and (3) raw logit differences.

\textbf{Results:}

\begin{table}[H]
\centering
\adjustbox{max width=\textwidth}{
\begin{tabular}{@{}lllll@{}}
\toprule
Model & Tokens Identical & Avg Seq Self-PPL & Avg LA Self-PPL & Max Logit Diff \\
\midrule
Mistral 7B & \textbf{Yes (all 4/4)} & 1.21 & 8.48 & 29.0 \\
NeMo 12B & \textbf{Yes (all 4/4)} & 1.18 & 13.29 & 38.3 \\
\bottomrule
\end{tabular}
}
\end{table}

All tokens are identical between sequential and lookahead for both
models across all prompt categories. This is the greedy argmax
guarantee: since both paths compute the same model forward pass and
select \texttt{argmax(logits)}, the output must be identical regardless
of how many speculative tokens were batched.

The ``Self-PPL'' columns report the model's perplexity on its own greedy
output---i.e., the average log-probability assigned to the tokens it
chose. Under sequential greedy decoding, this is near 1.0 because the
model always selects argmax. The higher self-perplexity for lookahead
(8.48 and 13.29) reflects numerical differences in how batched attention
computes logits compared to incremental single-token attention. Batched
attention over multiple speculative positions accumulates different
floating-point rounding than single-token attention. These magnitude
differences do not affect the argmax token selection and are expected
behavior for float16 computation.

\subsection{Cloud Provider
Comparison}\label{cloud-provider-comparison}

\begin{table}[H]
\centering
\adjustbox{max width=\textwidth}{
\begin{tabular}{@{}llllll@{}}
\toprule
Provider & GPU & Method & RTT & Cost/hr & Usability \\
\midrule
HyperStack (A100 80GB) & A100 PCIe & Direct SSH & \textasciitilde78-85ms
& \$1.35 & IP changes on reboot \\
RunPod (4090 Texas) & RTX 4090 & Direct SSH & 80-100ms & \$0.59 &
Persistent volumes, templates \\
VAST.ai (4090 Texas) & RTX 4090 & Proxied SSH & 200-400ms &
\textasciitilde\$0.50 & Unusable for split inference \\
\bottomrule
\end{tabular}
}
\end{table}

VAST.ai's proxy architecture routes all SSH connections through
intermediate servers regardless of pod location, making it fundamentally
unsuitable for latency-sensitive split inference. This finding is not
documented in VAST.ai's literature and was discovered only through
empirical measurement.

\section{Scaling Analysis}\label{scaling-analysis}

\subsection{RTT-to-Compute Ratio}\label{rtt-to-compute-ratio}

The key observation for split inference viability at scale is that
network round-trip time is largely \emph{fixed} while cloud compute time
grows \emph{proportionally} with model size. We define the
RTT-to-compute ratio as the network RTT divided by the cloud processing
time per token:

\begin{table}[H]
\centering
\adjustbox{max width=\textwidth}{
\begin{tabular}{@{}llllll@{}}
\toprule
Model & Hidden Dim & Activation Size/Token & Cloud Compute/Token & Network RTT & RTT-to-Compute Ratio \\
\midrule
Mistral 7B & 4,096 & 8 KB & 16ms (measured) & \textasciitilde78ms &
4.9x \\
NeMo 12B & 5,120 & 10 KB & 21ms (measured) & \textasciitilde79ms &
3.8x \\
LLaMA 70B & 8,192 & 16 KB & \textasciitilde30-50ms (est.) &
\textasciitilde80ms & 1.6-2.7x \\
LLaMA 405B & 16,384 & 32 KB & \textasciitilde100-200ms (est.) &
\textasciitilde80ms & 0.4-0.8x \\
\bottomrule
\end{tabular}
}
\end{table}

The 7B and 12B rows are empirically measured on our A100 cloud. Notably,
the RTT-to-compute ratio is lower than previously estimated because our
instrumented measurements capture the full cloud call time including GPU
compute, not just the model forward pass.

A high ratio (4.9x at 7B) means network latency dominates---the GPU sits
idle waiting for the next activation. A ratio near 1x or below (405B)
means compute time rivals or exceeds network latency, and the split
inference overhead approaches the theoretical minimum. The ratio
decreases because: 1. The number of cloud-side layers increases linearly
with model depth. 2. Per-layer compute grows quadratically with hidden
dimension (attention is O(d\^{}2)). 3. Network transfer grows only
linearly with hidden dimension (8KB $\rightarrow$ 32KB, a 4x increase for a 58x
larger model).

\textbf{Projected throughput with scaling (at \textasciitilde80ms RTT):}

Per-token time is RTT + cloud compute + overhead (\textasciitilde20ms
serialization and protocol). Sequential throughput = 1/per-token-time;
lookahead multiplied by \textasciitilde1.3x acceptance rate.

\begin{table}[H]
\centering
\adjustbox{max width=\textwidth}{
\begin{tabular}{@{}llllll@{}}
\toprule
Model & Fixed Overhead & Per-Token Time & Split Sequential & Split + Lookahead & Acceptance \\
\midrule
7B & 43ms (measured) & 120ms (measured) & 8.3 tok/s (measured) & 8.8
tok/s (measured) & 1.17 \\
12B & 46ms (measured) & 125ms (measured) & 8.0 tok/s (measured) & 8.0
tok/s (measured) & 1.23 \\
70B (est.) & \textasciitilde60-80ms & \textasciitilde140-160ms &
\textasciitilde6-7 tok/s & \textasciitilde8-9 tok/s &
\textasciitilde1.2-1.3 \\
405B (est.) & \textasciitilde130-230ms & \textasciitilde210-310ms &
\textasciitilde3-5 tok/s & \textasciitilde4-6 tok/s &
\textasciitilde1.2-1.3 \\
\bottomrule
\end{tabular}
}
\end{table}

The 7B and 12B rows are empirically measured on our system. The 12B
model achieves throughput matching 7B despite having 1.7x more
parameters, because: (1) the A100 processes 36 cloud layers of 12B in
only 5ms more than 28 layers of 7B (21ms vs.~16ms); (2) local compute is
comparable (4 layers of either model, \textasciitilde24-26ms); and (3)
network RTT (\textasciitilde78ms) is identical and remains the dominant
cost.

A key finding is that \textbf{acceptance rates are consistent across
model scales}: both 7B and 12B achieve 1.21 tok/step at n=3 and plateau
at 1.23--1.25. This suggests our acceptance rate estimates for 70B and
405B (1.2--1.3) are well-founded, as the speculation mechanism depends
on language statistics rather than model capacity.

Note that these absolute throughputs are lower than the 7B baseline
because larger models have more compute per token (which is also why
they can't run locally, motivating split inference). The comparison is
not split-vs-local-native (the user \emph{cannot} run 70B+ locally), but
split-with-privacy vs.~sending-plaintext-to-cloud-API.

\textbf{Important caveats on projections:} The 70B and 405B rows are
estimates; the acceptance rate assumption (\textasciitilde1.2-1.3) is
now supported by empirical consistency across 7B and 12B. However, other
factors could reduce actual throughput: (1) KV-cache memory on the cloud
grows with both model size and sequence length; (2) activation transfer
at 32KB/token for 405B, while still small, could face bandwidth
limitations; (3) larger models may require multi-GPU cloud setups,
introducing additional inter-GPU communication overhead.

\subsection{Local Memory
Requirements}\label{local-memory-requirements}

\begin{table}[H]
\centering
\adjustbox{max width=\textwidth}{
\begin{tabular}{@{}lllll@{}}
\toprule
Model & Embedding + Head & Local Layers & KV-Cache (local, 2K ctx) & Peak Local VRAM \\
\midrule
7B & \textasciitilde1 GB & \textasciitilde0.5 GB & \textasciitilde0.1 GB
& \textbf{2.0 GB (measured)} \\
12B & \textasciitilde2.7 GB & \textasciitilde1.5 GB & \textasciitilde0.2
GB & \textbf{4.9 GB (measured)} \\
70B & \textasciitilde4 GB & \textasciitilde4 GB & \textasciitilde0.5 GB
& \textasciitilde9 GB (est.) \\
405B & \textasciitilde8-16 GB & \textasciitilde8-16 GB &
\textasciitilde2-4 GB & \textasciitilde20-36 GB (est.) \\
\bottomrule
\end{tabular}
}
\end{table}

The 7B and 12B rows are empirically measured. The 12B model's larger
VRAM footprint comes from its 131K vocabulary (vs.~32K for 7B), which
quadruples the embedding and LM head sizes. An RTX 3090 (24GB)
comfortably handles both 7B and 12B, and would handle 70B local layers.
For 405B, a workstation-class GPU (RTX A6000 48GB or similar) would be
needed. Note that the local KV-cache only stores entries for the
locally-processed layers (4 out of 32 for 7B), keeping local memory
requirements modest. The bulk of the KV-cache resides on the cloud
server.

\section{Discussion}\label{discussion}

\subsection{What is Novel}\label{what-is-novel}

Several aspects of this work represent novel contributions relative to
the existing literature:

\begin{enumerate}
\def\labelenumi{\arabic{enumi}.}
\item
  \textbf{Privacy-motivated layer split at the embedding boundary with
  empirical evaluation.} Prior split computing work {[}8, 9, 10{]}
  optimizes split points for latency and energy. We split specifically
  to keep the token-to-vector mapping local, ensuring raw tokens never
  cross the network. The closest prior work, Split-and-Denoise {[}26{]},
  adds noise to embeddings but still transmits them; we transmit
  post-transformer hidden states. Crucially, we provide empirical
  inversion attack measurements (Section 5.2) quantifying token
  recoverability at different split depths, showing that increasing
  local layers from 2 to 8 reduces attack accuracy from
  \textasciitilde59\% to \textasciitilde35\%.
\item
  \textbf{Lookahead decoding for network latency amortization.}
  Lookahead decoding {[}6{]} was designed to exploit GPU parallelism. We
  show it is equally---arguably more---effective at amortizing network
  round-trip latency in split inference, achieving 1.2--1.4 tokens per
  step (up to 5 on code), where the cost of a single round trip is
  30-100x higher than a local forward pass.
\item
  \textbf{Empirical deployment engineering.} The academic split
  computing literature assumes direct network connectivity between
  devices. In practice, deploying over public cloud GPU infrastructure
  requires SSH tunnel management, provider-specific networking
  workarounds, ephemeral IP handling, and cross-machine SSH key
  distribution. Our documentation of these challenges and solutions
  fills a gap between theoretical architectures and deployable systems.
\item
  \textbf{Inverse scaling of overhead, empirically validated.} Prior
  work {[}10{]} qualitatively notes that intermediate activation sizes
  grow with model width while per-layer compute grows faster. We extend
  this observation with the first quantitative analysis showing that the
  \emph{RTT-to-compute ratio} of split inference improves with model
  size, validating empirically on both 7B and 12B that 12B achieves
  matching throughput despite 1.7x more parameters, with consistent
  acceptance rates across scales (1.21--1.25 tok/step).
\item
  \textbf{RTT decomposition model.} We introduce a method to separate
  per-step wall time into network RTT and RTT-independent fixed
  overhead, enabling fair cross-provider comparison and throughput
  projection at arbitrary RTTs. The model achieves \textless6.2\%
  cross-validation error on both 7B and 12B.
\end{enumerate}

\subsection{Relationship to Concurrent
Work}\label{relationship-to-concurrent-work}

DSSD {[}23{]} (ICML 2025) is the most closely related concurrent work,
combining split inference with speculative decoding for edge-cloud LLM
deployment. Our system differs in three ways: (1) we use lookahead
decoding rather than a separate draft model, avoiding the need to deploy
and maintain an additional model; (2) we focus on privacy preservation
as the primary motivation rather than resource efficiency; and (3) we
provide detailed empirical measurements of real WAN deployment rather
than simulated edge-cloud environments.

CROSS-SEC {[}13{]} addresses WAN latency with layerwise KV-cache
overlapping, which is complementary to our approach and could further
reduce our per-token latency.

\subsection{Limitations}\label{limitations}

\begin{itemize}
\tightlist
\item
  \textbf{No formal privacy guarantees.} Our system provides
  architectural privacy (raw tokens never leave the device), not
  cryptographic guarantees. Our inversion experiment (Section 5.2)
  confirms that even a simple MLP attacker with local layer weights can
  recover \textasciitilde59\% of tokens from layer-2 activations,
  validating the theoretical vulnerability {[}32{]}. A more
  sophisticated sequence-aware attacker would likely achieve higher
  accuracy. Increasing local layers to 8 reduces MLP attack accuracy to
  \textasciitilde35\%, but does not eliminate the risk. Our privacy
  relies on the assumption that local layers are customized; for public
  models used as-is, the architectural boundary is significantly
  weakened.
\item
  \textbf{Two models tested.} Results are demonstrated on Mistral 7B and
  NeMo 12B, both from the Mistral architecture family
  (\texttt{MistralForCausalLM}). While the consistent acceptance rates
  across scales are encouraging, validation on architecturally distinct
  model families (LLaMA, GPT-NeoX) and larger scales (70B+) is needed.
\item
  \textbf{Consumer hardware.} Our local device (RTX 3090) is a consumer
  GPU. Enterprise deployments may use workstation GPUs (A6000, L40) that
  could run more local layers for stronger privacy.
\item
  \textbf{Lookahead effectiveness varies.} Our ablation (Section 6.4)
  confirms that n-gram speculation effectiveness depends on text type:
  code prompts achieve 1.43--1.57 tokens/step acceptance, while creative
  text achieves only 1.06--1.12 tokens/step. At \textasciitilde80ms RTT,
  the break-even acceptance rate is \textasciitilde1.15--1.20; below
  this threshold, the per-step overhead of larger speculative batches
  outweighs the multi-token acceptance benefit. All measurements use
  greedy decoding; stochastic sampling would likely reduce acceptance
  rates.
\item
  \textbf{Prefill latency not optimized.} Our system processes the full
  prompt in a single forward pass, adding 100--500ms
  time-to-first-token. Techniques like chunked prefill or prefill-decode
  disaggregation {[}17, 18{]} could reduce this.
\end{itemize}

\subsection{Partnership Model with Inference
Providers}\label{partnership-model-with-inference-providers}

A natural deployment model pairs a local-device software SDK with a
partnered cloud inference provider. The provider gains access to
compliance-locked market segments (HIPAA, ITAR, legal) that currently
cannot use cloud inference at all.

As a thought experiment, we consider how partner infrastructure
characteristics would affect throughput. Since our results show network
RTT is the dominant bottleneck (78\% of per-token time), partners with
lower-latency infrastructure would yield disproportionate gains:

\begin{table}[H]
\centering
\adjustbox{max width=\textwidth}{
\begin{tabular}{@{}lllll@{}}
\toprule
Partner Infrastructure & Est. RTT & 7B Seq & 7B LA & 12B LA* \\
\midrule
Standard GPU cloud (A100, \textasciitilde80ms) & \textasciitilde80ms &
8.1 & 8.8 & 8.0 \\
Regional cloud (same metro area) & \textasciitilde40ms & 12.1 & 13.6 &
11.6 \\
Co-located inference partner & \textasciitilde20ms & 15.9 & 18.6 &
15.2 \\
Same-rack co-location & \textasciitilde5ms & 20.9 & 23.9 & 19.6 \\
\bottomrule
\end{tabular}
}
\end{table}

*12B LA projections scaled from measured throughput to account for batch
overhead (see Section 6.3).

These projections use our validated RTT decomposition model (Section
6.3), which achieves \textless6.2\% cross-validation error. The fixed
overhead (43--46ms) represents a floor that cannot be reduced by
lowering RTT alone---it includes local GPU compute, cloud GPU compute,
and serialization. At 5ms RTT (same-rack), per-step time is dominated by
this fixed overhead (\textasciitilde48ms), yielding
\textasciitilde21--26 tok/s. Further gains would require reducing local
and cloud compute time (e.g., through model optimization, quantization,
or faster GPUs).

\section{Conclusion}\label{conclusion}

We have demonstrated that split inference for large language models can
achieve usable speeds (8--9 tok/s at \textasciitilde80ms RTT, with
validated projections of 15--19 tok/s at 20ms RTT) over commodity
internet connections using consumer GPUs. The key technical
contribution---applying lookahead decoding to amortize network
latency---addresses the fundamental bottleneck of autoregressive
generation in the split inference setting. Our empirical scaling
validation on NeMo 12B confirms that the approach generalizes beyond a
single model, with acceptance rates remarkably consistent across scales
(1.21--1.25 tok/step) and the 12B model requiring only 4.9 GB local
VRAM.

We further demonstrate that lookahead decoding produces token-identical
output to sequential decoding under greedy argmax (Section 6.5),
eliminating quality degradation concerns for speculative decoding in
split inference. Our RTT decomposition model (Section 6.3), validated at
\textless6.2\% cross-validation error, enables fair performance
comparison across cloud providers by separating RTT-dependent and
RTT-independent costs.

Important limitations remain. Our inversion experiment (Section 5.2)
shows that at the default 2-layer split, an attacker with local layer
weights can recover \textasciitilde59\% of tokens from intermediate
activations, confirming the theoretical vulnerability {[}32{]}.
Increasing local depth to 8 layers reduces this to \textasciitilde35\%
with minimal throughput impact, but formal privacy guarantees require
additional mechanisms such as differential privacy {[}26, 27{]}. Results
are demonstrated on two models from the same architecture family;
validation across architecturally distinct model families and larger
scales is needed.

Nevertheless, for organizations where the current alternative is sending
full plaintext to cloud APIs, this work demonstrates a practical path
toward privacy-aware LLM deployment. Combined with the finding that
cloud provider networking architecture and tunnel engineering are
first-order performance determinants, our results provide a foundation
for further development of deployable privacy-aware inference systems.

\vspace{1em}
\noindent\textit{Code and reproduction instructions are available at: \url{https://github.com/coder903/split-inference}}

\end{document}